\documentclass[aps,prl,preprint,groupedaddress]{revtex4-1}
\usepackage{amsmath}

\newcommand{\bra}[1]{\big\langle \, #1\,\big\vert}
\newcommand{\ket}[1]{\big\vert\, #1\,\big\rangle}

\newcommand{\avr}[2]{\left\langle \, #1\,\right\rangle_{#2}}

\begin{document}

%\preprint{}

\title{Quantum Ansible: Telegraph by Cloning a Known Pure State}


\author{N. Nikitin}
%\thanks{}
\affiliation{Lomonosov Moscow State University, Department of Physics, Russia}
\affiliation{Lomonosov Moscow State University, Skobeltsyn Institute of Nuclear Physics, Russia}
\affiliation{Institute for Theoretical and Experimental Physics, Russia}

\author{K. Toms}
\affiliation{University of New Mexico, USA}


\date{\today}

\begin{abstract}
In the framework of non-relativistic quantum mechanics we present a proposal 
of a gedunken experimental setup of a quantum system allowing information 
exchange. We  discuss the compatibility of the procedure with a few no-go 
theorems.
\end{abstract}

%\keywords{}

\maketitle

%\section{}
%\subsection{}
%\subsubsection{}

\section{Introduction}
\label{intro}

In the fundamental work by Einstein, Podolsky, and Rosen \cite{Einstein:1935rr} 
it was shown for the first time that if two quantum objects are in an entangled 
state then any measurement performed on one of them instantly affects the other.
Starting then many attempts were made to use this quantum property for 
superluminal communication between two macroscopic objects. However already in
the work by Furry \cite{PhysRev.49.393} it was shown that in absence of a 
classical channel the communication is impossible due to the pure probabilistic
nature of the outcome of quantum measurement by a macroscopic device. Later this 
conclusion was studied and generalized in many papers (e.g. \cite{eberhard} and 
\cite{PhysRevLett.104.140401}). The statement that quantum mechanics is non-local 
for micro-objects and is local for macroscopic objects is known as 
``no-signaling condition'' \cite{PRboxes:1984}.

In \cite{Herbert1982} an elegant attempt was made to use the quantum correlation
together with cloning of an arbitrary pure state in order to establish a 
superluminal communication. The procedure of cloning of an arbitrary pure state
may be written as
\begin{equation}
\label{clone-procedure}
\ket{\psi}\,\ket{0}\, \to\, \ket{\psi}\,\ket{\psi},
\end{equation}
where $\ket{0}$ -- is a supplementary known state. 
Quite fast it was shown \cite{Wootters1982}, \cite{DIEKS198227}, that the 
procedure (\ref{clone-procedure}) is not compatible with one of the basic 
postulates of quantum mechanics -- the superposition principle. This fact is now
known as ``no-cloning theorem''.

Any known pure state $\ket{\psi}$ can be cloned \cite{Park1970}. The cloning
is performed using a unitary operator $\hat U$, which is specially selected for 
the given vector $\ket{\psi}$ and a known supplementary vector $\ket{0}$.
The procedure of cloning of a known pure state may be written as: 
\begin{equation}
\label{clone-U}
\hat U\, \Big ( \ket{\psi} \otimes \ket {0}\Big )\, =\,\ket{\psi} \otimes \ket{\psi}.
\end{equation}
Using the same operator $\hat U$ and the same supplementary vector $\ket{0}$ 
it is possible to clone any pure state that is orthogonal to a known state 
$\ket{\psi}$. This important fact will be used in the present paper. 

A theorem on impossibility of broadcasting of an arbitrary mixed state 
(no-broadcast theorem) was proved much later \cite{PhysRevLett.76.2818}.
The exact broadcasting of an arbitrary mixed state is not possible, but the
approximate broadcasting can be performed \cite{PhysRevA.54.1844}.
The exact broadcasting of a known mixed state is always possible.

The word ``ansible" first appeared in a 1966 novel by Ursula K. Le Guin, 
referring to a fictional superluminal communication device.

\section{Gedanken experimental procedure}
\label{sec:1}

In this section we will modify the procedure described in \cite{DIEKS198227},
and show that this modification may lead to an interesting conclusions with no
contradictions to \cite{eberhard}, \cite{PhysRevLett.104.140401}, 
\cite{Wootters1982}, and \cite{DIEKS198227}.

Two experimenters, Alice and Bob, are based on Earth. They agree to select
a spatial Cartesian coordinate system $(x,\, y,\, z)$. The axis $y$ is chosen
to point towards $\alpha$ Centauri. Both experimenters have identical 
Stern--Gerlach devices that allow them to measure a spin projection of a
charged fermion onto any spatial direction. In addition Alice and Bob agree 
that if Alice will obtain a fermion with any spin projection along the $z$ axis 
it means ``0'', while any spin projection along the $x$ axis means ``1''.
Alice and Bob agreed not to perform any spin measurements along any other axes.

Bob then travels to $\alpha$ Centauri. Let us suppose that exactly in the middle
of the way there is a source of spin-correlated charged fermions. Let it be two
electrons in the spin-singlet Bell state
\begin{equation}
	\label{Psi-}
	\ket{\Psi^-}\, =\,\frac{1}{\sqrt{2}}\,
	\Big (
	\ket{+^{(A)}_{\vec n}}\otimes\ket{-^{(B)}_{\vec n}}\, -\, \ket{-^{(A)}_{\vec n}}\otimes\ket{+^{(B)}_{\vec n}}
	\Big ),
\end{equation}
where $\ket{\pm^{(i)}}_{\vec n}$ -- state of the electron with a spin projection
$s^{(i)}_{\vec n} = \pm 1/2$ onto spatial axis set by unitary vector $\vec n$. 
The state $\ket{\Psi^-}$ has a unique feature: spin anticorrelation holds along
any direction. The source is supposed to be well collimated and emits electrons
only along the $y$ axis. Without loss of generality suppose that the electron
``$A$'' propagates towards Earth, while the electron ``$B$'' propagates towards 
$\alpha$ Centauri. Both electrons reach their destinations simultaneously.

It is necessary to mention the spatial localization of fermions in 
Eq.~(\ref{Psi-}). Both fermions may be considered as narrow wave packets with 
the wave functions in the coordinate representation 
$\varphi^{(\, A)}({\vec r},\, t)$ and $\varphi^{(\, B)}({\vec r},\, t)$ such as
\begin{equation}
\label{perekritie}
\int\limits_{V_3} d {\vec r}\, \varphi^{(\, A)\, *}({\vec r},\, t)\, \varphi^{(\, B)}({\vec r}, \, t)\, \approx\, 0
\end{equation}
for any time $t$. In non-relativistic quantum mechanics the position and 
spin spaces are separated. So Eqs. (\ref{Psi-}) and (\ref{perekritie}) 
are always compatible. In quantum field theory the question of spatial
localization of two fermions is more complicated and is beyond the scope 
of this paper.

Bob arrives to $\alpha$ Centauri and would like to communicate with Alice.

In order to send ``0'' Bob measures the spin projection of his electron ``B''
onto the $z$ axis. Let the result of the measurement be $s^{(B)}_z = -1/2$.
Immediately after the Bob's measurement state (\ref{Psi-}) is reduced to 
$\ket{+^{(A)}_z}\otimes\ket{-^{(B)}_z}$, and Alice has the state 
$\ket{+^{(A)}_z}$.

The problem is that Alice does not know what state she has after Bob's 
measurement. In order to find the polarization of the obtained unknown state
Alice needs to prepare an ensemble of identical states. But no--cloning theorem
\cite{Wootters1982}, \cite{DIEKS198227} forbids such procedure.

Now Alice would like to apply the CNOT operator to {\it any} obtained state
$\ket{?^{\, (A)}}$. As the supplementary vector $\ket{0}$ she always chooses 
$\ket{+_z}$. In the basis
\begin{equation}
\label{basis}
\ket{+_z}\, =\,
\left (
\begin{matrix}
1 \\
0
\end{matrix}
\right ), \quad
\ket{-_z}\, =\,
\left (
\begin{matrix}
0 \\
1
\end{matrix}
\right )
\end{equation}
the CNOT operator \cite{Ucnot} may be written as 
\begin{eqnarray}
\label{cnot}
\hat U_{\textrm{CNOT}}\, =\,
\left (
\begin{matrix}
1 & 0 & 0 & 0 \\
0 & 1 & 0 & 0 \\
0 & 0 & 0 & 1 \\
0 & 0 & 1 & 0 
\end{matrix}
\right )
\end{eqnarray}
%\vspace{2pt}
Then for the state $\ket{+^{(A)}_z}$ one may write 
\begin{equation}
\label{++}
\hat U_{\textrm{CNOT}}\, \Big ( \ket{+^{(A)}_z} \otimes \ket {+_z}\Big )\, =\, \ket {+_z} \otimes \ket{+_z}.
\end{equation}
The outcome of the procedure (\ref{++}) is two identical states 
$\ket {+_z}$. Any of them, or even both, may be again subjected to the operator
CNOT. And so on. As the result Alice obtains an ensemble of identical states
$\ket {+_z}$. This ensemble can be divided in two sub-ensembles. For the first
one Alice may measure the average spin projection onto the $x$ axis. According
to the rules of quantum mechanics $\avr{S_x}{}=0$. For the second one Alice
may measure the average spin projection onto the $z$ axis. In this case 
$\avr{S_z}{}= +1/2$ will be found.

If Bob measures $s^{(B)}_z = +1/2$, then Alice immediately has 
$\ket{-^{(A)}_z}$. Application of the operator (\ref{cnot}) to this state gives 
\begin{equation}
\label{-+}
\hat U_{\textrm{CNOT}}\, \Big ( \ket{-^{(A)}_z} \otimes \ket {+_z}\Big )\, =\, \ket {-_z} \otimes \ket{-_z},
\end{equation}
because the state $\ket{-^{(A)}_z}$ is orthogonal to the state 
$\ket{+^{(A)}_z}$.
Continious application of the procedure (\ref{-+}) will allow Alice to get an 
ensemble of states $\ket {-_z}$. After dividing this ensemble to sub-ensembles 
and measurements of the average spin projections onto the $x$ and $z$ axes
Alice will obtain the following results: $\avr{S_x}{}=0$ and $\avr{S_z}{}= -1/2$.

Now consider the situation when Bob would like to broadcast ``$1$''. In this 
case Bob should measure the spin projection of his electron ``$B$'' onto the $x$
axis. Suppose the result of the measurement is $s^{(B)}_x = -1/2$. Then
the (\ref{Psi-}) state should immediately collapse to 
$\ket{+^{(A)}_x}\otimes\ket{-^{(B)}_x}$. I.e. Alice will instantly have 
$\ket{+^{(A)}_x}$, which she interprets as $\ket{?^{\, (A)}}$. Alice then 
apply the linear operator (\ref{cnot}) to this state. As 
\begin{eqnarray}
\label{x-basis}
\ket{+_x}\, =\,\frac{1}{\sqrt{2}}\,
\left (
\begin{matrix}
1 \\
1
\end{matrix}
\right )\, =\,
\frac{1}{\sqrt{2}}\,
\Big (
\ket {+_z}\, +\, \ket {-_z}
\Big),
\end{eqnarray}
then according to (\ref{++}) and (\ref{-+}), (\ref{x-basis}) and due to 
the linearity of $\hat U_{\textrm{CNOT}}$ we can write
\begin{widetext}
\begin{eqnarray}
\label{x++}
&&\hat U_{\textrm{CNOT}}\, \Big ( \ket{+^{(A)}_x} \otimes \ket {+_z}\Big )\, =
\nonumber \\
&=&\frac{1}{\sqrt{2}}
\left [
\hat U_{\textrm{CNOT}}\, \Big ( \ket{+^{(A)}_z} \otimes \ket {+_z}\Big ) +
\hat U_{\textrm{CNOT}}\, \Big ( \ket{-^{(A)}_z} \otimes \ket {+_z}\Big )
\right ] = 
 \\
&=&
\frac{1}{\sqrt{2}}\,
\Big ( \ket {+_z^{(1)}} \otimes \ket{+_z^{(2)}}\, +\, \ket {-_z^{(1)}} \otimes \ket{-_z^{(2)}} \Big )\, =\, \ket{\Phi^+}.
\nonumber
\end{eqnarray}
\end{widetext}
In Eq. (\ref{x++}) the indices ``$1$'' and ``$2$'' denote the states of each 
of the obtained particles. Note that unlike (\ref{++}) and (\ref{-+}), where
the final states were factorized, the final state in the Eq. (\ref{x++}) 
is the entangled Bell state $\ket{\Phi^+}$. This means that each of the 
particles ``$1$'' and ``$2$'' is in a fully unpolarized states, that are 
described by the following density matrices:
\begin{equation}
\label{rho1}
\hat\rho^{(1)} =\,\textrm{Tr}_2\,\left (\ket{\Phi^+}\bra{\Phi^+} \right )\, =\,\frac{1}{2}\,\hat 1^{(1)}
\end{equation}
and
\begin{equation}
\label{rho2}
\hat\rho^{(2)} =\,\textrm{Tr}_1\,\left (\ket{\Phi^+}\bra{\Phi^+} \right )\, =\,\frac{1}{2}\,\hat 1^{(2)},
\end{equation}
where $\hat 1^{(i)}$ is a unitary matrix $2 \times 2$ in two-dimensional 
Hilbert space ${\cal H}^{(i)}$ states of the particle $i$. 
While obtaining (\ref{rho1}) and (\ref{rho2}) we used the orthogonal 
decomposition of a unitary operator
$$
\hat 1^{(i)} = \hat P^{(i)}_{+\, z} + \hat P^{(i)}_{-\, z},
$$
where $\hat P^{(i)}_{\pm\, z} = \ket{\pm^{(i)}_{\, z}}\bra{\pm^{(i)}_{\, z}}$ 
are the projectors onto the corresponding pure states. 

So both particles in the pure state $\ket{\Phi^+}$ are described by the 
density matrices of the same kind. Alice may subject any of them to the operator
$\hat U_{\textrm{CNOT}}$. Let is be particle ``$1$''. Then
\begin{widetext}
\begin{eqnarray}
\label{rho-clon}
&&\hat U_{\textrm{CNOT}}\,
\left (
\frac{1}{2}\,\hat 1^{(1)}\,\otimes\,\ket{+_z}\bra{+_z}
\right )\,
\hat U_{\textrm{CNOT}}^{\dagger}\, =
\nonumber \\
&=&
\frac{1}{2}\, 
\hat U_{\textrm{CNOT}}\,
\Big (
\ket{+^{(1)}_z}\,\otimes\,\ket{+_z}\bra{+_z}\otimes\bra{+^{(1)}_z}
\Big )\,
\hat U_{\textrm{CNOT}}^{\dagger}\, +
\\
&+&
\frac{1}{2}\, 
\hat U_{\textrm{CNOT}}\,
\Big (
\ket{-^{(1)}_z}\,\otimes\,\ket{+_z}\bra{+_z}\otimes\bra{-^{(1)}_z}
\Big )\,
\hat U_{\textrm{CNOT}}^{\dagger}\, =
\nonumber \\
&=&
\frac{1}{2}\, 
\left (
\hat P^{(1)}_{+ \, z} \otimes \hat P^{(3)}_{+ \, z}\, +\,
\hat P^{(1)}_{- \, z} \otimes \hat P^{(3)}_{- \, z}\
\right )\, =\, \hat\rho.
\nonumber
\end{eqnarray}
\end{widetext}
The result of the procedure (\ref{rho-clon}) is a separable state $\hat\rho$ of 
two particles ``$1$'' and ``$3$''. Easy to see that the particle ``$1$'' is 
still in an unpolarized mixed state 
$\displaystyle\frac{1}{2}\,\hat 1^{(1)} = \frac{1}{2}\, 
\left (\hat P^{(1)}_{+ \, z} +\, \hat P^{(1)}_{- \, z} \right )$, 
while the particle ``$3$'', which was created from a supplementary particle 
in the state $\ket{+_z}$, after the application of the operator 
$\hat U_{\textrm{CNOT}}$ went to an unpolarized state 
$\displaystyle\frac{1}{2}\,\hat 1^{(3)}$.

I.e. after (\ref{x++}) and (\ref{rho-clon}) Alice has three identical particles
``$1$'', ``$2$'', and ``$3$'', and each of them is in a fully unpolarized state
$\displaystyle\frac{1}{2}\,\hat 1$. It is obvious that the consecutive 
application of the procedure (\ref{rho-clon}) to these three particles will 
allow Alice to obtain an ensemble of identical particles, with every one of
them in the state $\displaystyle\frac{1}{2}\,\hat 1$. Then Alice divides this 
ensemble to two sub-ensembles. For the first ensemble Alice may measure 
the average spin projection onto the $x$ axis, which will result in 
$\avr{S_x}{}=0$.  For the second -- the average spin projection onto the $z$
axis. The particles are not polarized, hence $\avr{S_z}{}= 0$. 

Finally consider the situation when Bob got $s^{(B)}_x = +1/2$. In this case 
according to Eq.~(\ref{Psi-}) Alice will immediately have a pure state 
$\ket{-^{(A)}_x}$. We emphasize once again that Alice does not know what 
state she has. Takin into account that 
\begin{eqnarray}
\label{xx-basis}
\ket{-_x}\, =\,\frac{1}{\sqrt{2}}\,
\left (
\begin{matrix}
-1 \\
1
\end{matrix}
\right )\, =\, -\,
\frac{1}{\sqrt{2}}\,
\Big (
\ket {+_z}\, -\, \ket {-_z}
\Big),
\end{eqnarray}
we found the result of application of the operator (\ref{cnot}) to the state 
$\ket{-^{(A)}_x} \otimes \ket {+_z}$:
\begin{widetext}
\begin{eqnarray}
\label{x-+}
&&\hat U_{\textrm{CNOT}}\, \Big ( \ket{-^{(A)}_x} \otimes \ket {+_z}\Big )\, =
\\
&=& -\,\frac{1}{\sqrt{2}}\,
\Big ( \ket {+_z^{(1)}} \otimes \ket{+_z^{(2)}}\, -\, \ket {-_z^{(1)}} \otimes \ket{-_z^{(2)}} \Big )\, =\, -\, \ket{\Phi^-}.
\nonumber
\end{eqnarray}
\end{widetext}
In the entangled Bell state $\ket{\Phi^-}$ each of the particles ``$1$'' or 
``$2$'' is in the fully unpolarized state $\displaystyle\frac{1}{2}\,\hat 1$. 
Applying the transformation (\ref{rho-clon}) Alice may create an ensemble of 
identical particles in the unpolarized state $\displaystyle\frac{1}{2}\,\hat 1$. 
Further, like in the above cases, Alice divides this ensemble into two 
sub-ensembles. For the first the average spin is measured along the $x$ axis, 
for the second -- along the $y$ axis. Measurements result in 
$\avr{S_x}{}=\avr{S_z}{}=0$.

\section{Discussion}
\label{sec:2}

Alice subjects any unknown state $\ket{?^{(A)}}$ to the same procedure: 
application of the operator $\hat U_{\textrm{CNOT}}$ (\ref{cnot}) using the 
same supplementary vector $\ket{0}=\ket{+_z}$. Then the operator 
$\hat U_{\textrm{CNOT}}$ is applied again and again using $\ket{+_z}$.
Then the obtained particled are divided into two sub-ensembles. For the first 
one the average spin projection $\avr{S_x}{}$ is measured onto the $x$ axis, 
for the second one the average spin projection $\avr{S_z}{}$ -- onto the $z$ 
axis. During these measurements Alice and Bob do not have any means of 
classical communication.

If Bob performed any spin measurement along the $z$ axis then, as it was shown 
above, Alice, using her procedure, immediately obtains that 
$\avr{S_x}{}=0$, while $\avr{S_z}{}= \pm\, 1/2$. On the other hand, if Bob 
performed a spin measurement along the $x$ axis, then Alice will obtain  
$\avr{S_x}{}=0$ and $\avr{S_z}{}= 0$.

I.e. Alice may distinct locally on Earth, in absence of any classical 
communication, what measurement was performed by Bob at $\alpha$ Centauri. 
This opens a possibility to use the superluminal binary code for the 
information exchange. 

At first sight this statement contradicts to \cite{PhysRev.49.393} -- 
\cite{PhysRevLett.104.140401} and directly contradicts to the no-signalling 
condition~\cite{PRboxes:1984}. But this in not true, as during the derivations
of \cite{PhysRev.49.393} -- \cite{PhysRevLett.104.140401} it was not supposed
that Alice and Bob created an additional local correlation: an agreement
that the measurements be performed only along the two axes: $x$ and $y$.
It is obvious that without this condition Alice's measurement procedure
would be unsuccessful.

\section*{Conclusions}

In the current paper we present a description of quantum ansible: a 
superluminal binary telegraph that does not violate the no-signalling 
condition, does not contradict to the no-cloning theorem and does not 
contradict to the no-broadcast theorem. The procedure is considered in 
the framework of non-relativistic quantum mechanics. The question of the 
possibility of a similar procedure in the framework of quantum field theory 
remains open.

\section*{Acknowledgements}

We would like to thank C.~Aleister (Saint~Genis--Pouilly, France) for creating 
a warm and friendly working atmosphere for discussions between the authors.

The work was supported by grant 16-12-10280 of the Russian Science
Foundation. One of the authors (N.~Nikitin) expresses his gratitude for
this support.


\begin{thebibliography}{99}
\bibitem{Einstein:1935rr}
A.~Einstein, B.~Podolsky and N.~Rosen, ``Can quantum mechanical description of physical reality be considered complete?'', Phys.\ Rev.\  {\bf 47}, 777  (1935).
\bibitem{PhysRev.49.393} 
W.~H.~Furry, ``Note on the Quantum--Mechanical Theory of Measurement'', Phys.\ Rev.\  {\bf 49}, 393 (1936).
\bibitem{eberhard} 
P.~H.~Eberhard, ``Bell's Theorem and the Different Concepts of Locality'', Nuovo Cimento \textbf{46}B, 392 (1978).
\bibitem{PhysRevLett.104.140401} 
H.~Barnum, S.~Beigi, S.~Boixo, M.~B.~Elliott,  and S.~Wehner, ``Local Quantum Measurement and No-Signaling Imply Quantum Correlations'', Phys.\ Rev.\ Lett.\ {\bf 104}, 140401 (2010).
\bibitem{PRboxes:1984} 
S.~Popescu and D.~Rohrlich, ``Quantum nonlocality as an axiom'', Found.\ Phys. \textbf{24}, 379 (1994).
\bibitem{Herbert1982} 
N.~Herbert, ``FLASH -- A superluminal communicator based upon a new kind of quantum measurement'', Found.\ Phys.\ {\bf 12}, 1171 (1982).
\bibitem{Wootters1982} 
W.~K.~Wootters and W.~H.~Zurek, ``A single quantum cannot be cloned'', Nature \textbf{299}, 802 (1982). 
\bibitem{DIEKS198227} 
D.~Dieks, ``Communication by EPR devices'', Phys.\ Lett.\  A {bf 92}, 271 (1982).
\bibitem{Park1970}
 J.~L~Park, ``The concept of transition in quantum mechanics'', Found.\ Phys.\ {\bf 1}, 23 (1970).
\bibitem{PhysRevLett.76.2818} 
H.~Barnum, C.~M.~Caves, C.~A.~Fuchs, R.~Jozsa, and B.~Schumacher, ``Noncommuting Mixed States Cannot Be Broadcast'', Phys.\  Rev.\  Lett.\ {\bf 76}, 2818 (1996).
\bibitem{PhysRevA.54.1844}
V.~Buzek and M.~Hillery, ``Quantum copying: Beyond the no-cloning theorem'', Phys.\ Rev.\ A\ {\bf 54}, 1844 (1996).
\bibitem{Ucnot}
M.~A.~Nielsen, I.~L.~Chuang, ``Quantum Computation and Quantum Information: 10th Anniversary Edition'', Cambridge University Press, New York, NY, USA (2011).
\end{thebibliography}
\end{document}